\documentclass[journal]{IEEEtran}

\usepackage{cite}
\usepackage{amsmath,amssymb,amsfonts}
\usepackage{algorithmic}
\usepackage{graphicx}
\usepackage{textcomp}
\usepackage{xcolor}
\usepackage{styles/mathnotation}
\usepackage{booktabs}
\usepackage{multirow}
\usepackage{makecell}
\usepackage{stfloats}
\usepackage{placeins}
\usepackage{needspace}
\newcommand{\Nrx}{N_{\mathrm{rx}}}
\newcommand{\Ntx}{N_{\mathrm{tx}}}
\newcommand{\Nsc}{N_{\mathrm{sc}}}
\newcommand{\Nsym}{N_{\mathrm{sym}}}
\newcommand{\Nre}{N_{\mathrm{re}}}
\newcommand{\nce}{L_{\mathrm{ce}}}
\newcommand{\nsd}{L_{\mathrm{sd}}}
\newcommand{\hls}{\wh{\Hv}^{\mathrm{LS}}}
\newcommand{\hest}{\wh{\Hv}}
\newcommand{\Ldet}{\mathcal{L}_{\mathrm{det}}}
\newcommand{\Lceaux}{\mathcal{L}_{\mathrm{ce}}}
\newcommand{\Ltotal}{\mathcal{L}_{\mathrm{total}}}

\newcommand{\Nprb}{N_{\mathrm{prb}}}
\newcommand{\Wf}{W_{\mathrm{f}}}

\newcommand{\ourmethods}{SWAN-Rx}

\begin{document}

\title{SWAN-Rx: Efficient and Bandwidth-Scalable Neural Reception for MIMO-OFDM}
\author{Zhoubin~Kou,~\IEEEmembership{Graduate Student Member,~IEEE,}
and Cong~Shen,~\IEEEmembership{Senior Member,~IEEE}\thanks{A preliminary version of this work was accepted for presentation at the 2026 IEEE Global Communications Conference (GLOBECOM).}\thanks{Zhoubin Kou and Cong Shen are with the Department of Electrical and Computer Engineering, University of Virginia, Charlottesville, VA, USA (e-mail: \{zhoubin, cong\}@virginia.edu).}}
\maketitle

\begin{abstract}

We present a Scalable Windowed Axial-attention Neural Receiver (\ourmethods{}) for multiple-input multiple-output orthogonal frequency-division multiplexing (MIMO-OFDM) systems. The receiver improves reception accuracy and reduces forward-pass computation while supporting frozen-weight reuse across bandwidth allocations and channel conditions. It combines a common resource-element representation of received observations and pilot context with alternating time-axis and local frequency-axis attention. Shared parameters and relative positional information permit soft detection across allocation lengths without inference-time parameter updates. At fixed subcarrier spacing, model configuration, and symbol count, bounded frequency neighborhoods give linear growth of forward-pass computation with bandwidth. Evaluations compatible with the 5G New Radio physical layer reuse weights trained on four physical resource blocks (PRBs) at up to 273 PRBs and on unseen channels. At the training bandwidth, \ourmethods{} achieves lower pre- and post-decoding bit error rates than the implemented three-step neural receiver (NRX) and DeepRx-MIMO baselines, with 16.9\% and 39.7\% fewer parameters and 56.3\% and 35.7\% fewer forward-pass operations, respectively. The same weights achieve lower pre-decoding error rates than these baselines throughout the evaluated bandwidth-extrapolation and 16-PRB channel-transfer tests, with an estimated 0.4--0.9~dB signal-to-noise ratio reduction relative to NRX at a pre-decoding error rate of $10^{-2}$ in the 16-PRB tests.
\end{abstract}


\section{Introduction}
\label{sec:intro}
Neural receivers offer a learned approach to physical-layer processing in modern wireless systems, mapping received observations and pilot information to soft bit outputs for channel decoding \cite{honkala2021deeprx,korpi2021deeprx,cammerer2023neural}. Practical deployment in systems such as 5G New Radio (NR) requires compatibility with multiple prescribed transmission structures, resource configurations, and processing constraints. In mobile multiple-input multiple-output orthogonal frequency-division multiplexing (MIMO-OFDM) links, multipath propagation introduces frequency selectivity, while mobility causes the channel to evolve across OFDM symbols. Reliable reception therefore depends on processing observations across both time and frequency, accounting for channel evolution within the transmission block. Varying physical resource block (PRB) allocations and propagation conditions pose a significant challenge to neural receivers, further motivating reuse of learned weights across operating configurations and reducing (or eliminating) the need to re-train and maintain separate models. 

Existing neural receivers address configuration flexibility through structured input representations and shared processing~\cite{honkala2021deeprx,korpi2021deeprx,cammerer2023neural}. Shared convolutional kernels enable processing over resource grids of different sizes, while the neural receiver (NRX) of Cammerer et al.~\cite{cammerer2023neural} \space additionally incorporates graph-based processing across spatial streams and supports varying subcarrier counts without retraining.
Computational efficiency is another consideration for practical deployment. The real-time NRX study in~\cite{wiesmayr2024realtime} shows that meeting a latency budget can require reducing iteration depth, with a corresponding loss in reception performance. 

Attention-based architectures offer another approach to meeting these processing constraints. Inspired by axial attention for multidimensional data~\cite{ho2019axial}, Yellapragada et al.~\cite{yellapragada2026computationally} developed an axial-attention neural receiver for single-input multiple-output (SIMO) OFDM systems. By processing the time and frequency axes separately, their receiver achieves competitive reception performance with lower computational cost than the evaluated convolutional baselines \cite{honkala2021deeprx}. However, several aspects of this design limit its applicability across transmission configurations. Its input representation combines received observations with noise information but does not explicitly encode known pilot symbols or distinguish pilot positions from data positions~\cite{yellapragada2026computationally}. Its learned positional tensor is tied to the resource-grid dimensions, so direct reuse at larger PRB allocations requires adaptation of the positional representation. Moreover, attention over the entire frequency axis incurs a computational cost that grows quadratically with the number of subcarriers, limiting its scalability to wideband allocations. These limitations motivate an input representation that explicitly incorporates pilot context, together with attention processing that supports variable allocation lengths at a controlled computational cost.

In this work, we propose the Scalable Windowed Axial-attention Neural Receiver (SWAN-Rx), which is an end-to-end MIMO-OFDM receiver with a common resource-element representation of pilot and data observations, local axial processing, and relative positional information. Beyond detection, the axial structure facilitates inspection of attention patterns along the time and frequency axes and supports a multitask extension for joint channel estimation and signal detection. This extension produces explicit channel estimates alongside soft bit outputs, with stable channel-estimation accuracy across the evaluated bandwidths and channel conditions. The main contributions are as follows.

$\bullet$ We introduce a pilot-conditioned resource-element representation for Transformer-based MIMO-OFDM reception. By organizing known pilot information as context and explicitly distinguishing pilot and data symbols, the representation connects the wireless transmission structure to the receiver's attention processing and provides a common input format across different PRB allocations.

$\bullet$ We develop a windowed axial-attention architecture tailored to the time--frequency resource grid. Time-axis attention and local frequency-axis attention, combined with relative positional information, support frozen-weight processing across bandwidth allocations while ensuring linear scaling of forward-pass computation with bandwidth for a fixed model configuration and symbol count.

$\bullet$ Through extensive experiments, we demonstrate improved reception accuracy and computational efficiency under a common narrowband-training protocol. At 4 PRBs, \ourmethods{} achieves lower pre- and post-decoding BER than the implemented NRX and DeepRx-MIMO baselines, with fewer parameters and 56.3\% and 35.7\% fewer forward-pass operations, respectively. The same weights achieve lower pre-decoding BER throughout the evaluated bandwidth-extrapolation tests up to 273 PRBs and the 16-PRB channel-transfer tests. Architectural ablations and NRX depth controls further support the design and comparative results.

The rest of this paper is organized as follows. Section~\ref{sec:related} reviews related work. Section~\ref{sec:prelim} introduces the signal model, pilot-conditioned reception, and attention operations. Section~\ref{sec:method} presents the receiver architecture, bandwidth scaling, and training procedure. Section~\ref{sec:results} reports the experimental results, and Section~\ref{sec:conclusion} concludes the paper.

\section{Related Work}
\label{sec:related}

\subsection{Structured Neural Receivers for MIMO-OFDM}
\label{ssec:related_receivers}

Neural receiver architectures differ in how they combine learned processing with the structure of the communication system.
DeepWiPHY jointly addresses channel estimation, synchronization-impairment correction, and equalization for IEEE 802.11ax receivers~\cite{zhang2020deepwiphy}.
DeepRx uses a fully convolutional network to process pilot and data observations and generate soft bits for channel decoding~\cite{honkala2021deeprx}.
DeepRx MIMO extends this approach to spatial multiplexing by introducing either a maximal-ratio-combining transformation or a learned multiplicative transformation before convolutional detection~\cite{korpi2021deeprx}.
These designs illustrate the importance of the input representation and spatial processing when extending learned reception to MIMO systems.

Configuration flexibility is addressed explicitly by the CNN/GNN receiver of Cammerer et al.~\cite{cammerer2023neural}.
Convolutional processing exploits time-frequency correlations, while graph-based information exchange accommodates multiple users and spatial layers.
The receiver supports variable subcarrier counts without retraining, establishing bandwidth reuse as an existing capability of neural receivers.
Its real-time extension further studies the relationship between iteration depth, reception performance, and implementation latency~\cite{wiesmayr2024realtime}.
This line of work motivates evaluating the computational cost of a reusable receiver together with its detection accuracy.

EqDeepRx pursues efficient MIMO reception through a hybrid architecture~\cite{honkala2026eqdeeprx}.
It augments pilot-based channel estimation with a learned DenoiseNN, applies parallel linear minimum mean-square error (LMMSE) and regularized zero-forcing equalizers, and feeds their outputs to a shared DetectorNN operating separately on each spatial layer.
A compact DemapperNN then produces bit log-likelihood ratios (LLRs).
The explicit equalizers reduce the burden on neural spatial processing, while the shared per-layer detector primarily addresses scalability across spatial multiplexing orders.
Our evaluation instead examines reception accuracy and computational cost when weights trained at a narrow bandwidth are reused at wider unseen PRB allocations.

\subsection{Pilot-Conditioned Reception and In-Context Learning}
\label{ssec:related_icl}

Pilots can support receiver adaptation through either parameter estimation or conditioning of a pretrained model.
RCNet exploits the time-frequency structure of MIMO-OFDM and fits reservoir-computing readout parameters using limited online reference signals~\cite{zhou2021rcnet}.
In-context learning (ICL) instead uses contextual examples to determine the receiver output without inference-time parameter updates~\cite{zecchin2025context}.
Zecchin et al. formulate MIMO equalization using transmitted and received pilot pairs as context~\cite{zecchin2023mimoicl}.
Kunde et al. establish asymptotic optimality of softmax-attention estimators for a specified subclass of in-context estimation problems~\cite{kunde2025transformers}.
Fan et al. extend the available context through decision feedback, incorporating previously detected symbols to assist subsequent detection over block-fading channels~\cite{fan2024decision}.

Related OFDM studies incorporate additional receiver structure into this formulation.
Kou et al. use pilots within a coherence block as prompts for joint symbol detection and explicit channel-frequency-response estimation~\cite{kou2025multitask}.
Fu studies a physics-guided MIMO-OFDM ICL receiver for sparse and superimposed pilot configurations~\cite{fu2026deep}.
Our formulation organizes pilot context at the resource-element (RE) level, combining received observations, known symbols, and pilot-derived channel estimates with explicit pilot/data roles.

\subsection{Efficient Attention and Bandwidth Generalization}
\label{ssec:related_attention}

Attention-based receivers differ in both their tokenization and their processing scope.
SigT represents the signal received at each antenna as a token for Transformer-based OFDM reception~\cite{ren2022sigt}, whereas the present formulation organizes observations and pilot context by RE.
Channelformer combines an attention encoder with a residual convolutional decoder for OFDM channel estimation and investigates pruning and online training from receiver-available information~\cite{luan2023channelformer}.
Its estimation and online-training objectives complement the frozen-weight soft-reception setting studied here.

Axial attention factorizes interactions along the axes of multidimensional data~\cite{ho2019axial}.
Local attention offers a complementary means of controlling interaction cost: Longformer combines sliding-window attention with selected global interactions to obtain linear scaling with sequence length~\cite{beltagy2020longformer}.
For OFDM reception, Yellapragada et al. apply time-axis and frequency-axis attention to a two-dimensional resource grid~\cite{yellapragada2026computationally}.
This reduces the interaction cost of full-grid attention, but the frequency stage still forms all subcarrier pairs, and the learned positional tensor is indexed by grid location.
Kawai and Koodli use RE-level embeddings and nonoverlapping resource-grid tiles for attention-based reception, alongside channel-estimation and interpolation applications~\cite{kawai2025unified}.
Both approaches use RE-level tokens, while their nonoverlapping tiling differs from the sliding frequency neighborhoods used here to exchange information across adjacent PRBs.
CHEA~\cite{abdollahpour2026scalable} uses multi-resolution windowed attention for channel estimation with linear complexity and bandwidth reuse. Our work instead targets pilot-conditioned soft detection and evaluates frozen-weight extrapolation beyond the training bandwidth, whereas CHEA is evaluated within its training bandwidth range.

Positional representations also affect reuse across allocation lengths.
Rotary position embeddings encode relative displacement in attention interactions~\cite{su2024roformer}.
In wireless channel estimation, ReFLEX uses relative-frequency bias to reuse a channel-state information (CSI) denoiser at unseen resource-block lengths~\cite{zhang2026reflex}. For end-to-end soft detection, \ourmethods{} combines relative positional information with sliding frequency windows, without learned frequency parameters tied to the training allocation.

\section{Preliminaries}
\label{sec:prelim}

We first define the MIMO-OFDM receiver design problem and its pilot-conditioned formulation, and then introduce the attention operations used to process the resource grid.

\subsection{MIMO-OFDM Signal Model}
\label{ssec:prelim_sysmodel}

We consider a single-user uplink with $\Ntx$ transmit antennas and $\Nrx \geq \Ntx$ receive antennas, with one spatial stream per transmit antenna.
An allocation contains $\Nsym$ OFDM symbols and $\Nsc=12\Nprb$ active subcarriers, where $\Nprb$ is the number of allocated PRBs.
The resource grid is indexed by
$\mathcal{G}=\{0,\ldots,\Nsym-1\}\times\{0,\ldots,\Nsc-1\}$.
Under ideal synchronization and a frequency-domain model neglecting intercarrier and intersymbol interference, the receiver observation after cyclic-prefix removal and the fast Fourier transform at RE $(t,k)$ is
\begin{equation}
    \yv_{t,k}=\Hv_{t,k}\xv_{t,k}+\nv_{t,k},
    \label{eq:signal}
\end{equation}
where $\xv_{t,k}\in\Cb^{\Ntx}$ is the transmitted symbol vector, $\yv_{t,k}\in\Cb^{\Nrx}$ is the received vector, and $\Hv_{t,k}\in\Cb^{\Nrx\times\Ntx}$ is the channel matrix.
The additive noise satisfies $\nv_{t,k}\sim\mathcal{CN}(\mathbf{0},\sigma^2\Iv_{\Nrx})$.
The channel may vary across both OFDM symbols and subcarriers.

Let $\mathcal{G}_{\mathrm p}$ and $\mathcal{G}_{\mathrm d}$ denote the pilot and data RE sets within the active grid.
The transmit vectors at pilot REs are known to the receiver, including the zero entries associated with inactive streams under frequency-orthogonal pilot transmission.
At data REs, each transmitted entry belongs to a constellation $\mathcal{S}$ carrying $B_s=\log_2|\mathcal{S}|$ coded bits per symbol.
We vary $\Nprb$ while keeping the OFDM-symbol count, antenna configuration, and subcarrier spacing $\Delta f$ fixed.
Thus, a subcarrier-index displacement of $\delta$ always represents the physical frequency separation $\delta\Delta f$, independent of the total allocation length.

\subsection{Pilot-Conditioned Soft Reception}
\label{ssec:prelim_transformer}

Define the received grid $\mathcal{Y}=\{\yv_{t,k}:(t,k)\in\mathcal{G}\}$ and the available pilot context
\begin{equation}
    \mathcal{C}_{\mathrm p}
    =\{(t,k,\xv_{t,k},\yv_{t,k}):(t,k)\in\mathcal{G}_{\mathrm p}\}.
    \label{eq:pilot_context}
\end{equation}
A conventional receiver uses this context to estimate the channel, followed by equalization and soft demapping.
A learned soft receiver can instead approximate the mapping from the observations and context to coded-bit LLRs:
\begin{equation}
    \{\boldsymbol{\ell}_{t,k}\}_{(t,k)\in\mathcal{G}_{\mathrm d}}
    =\mathcal{F}_{\theta}(\mathcal{Y},\mathcal{C}_{\mathrm p}),
    \qquad \boldsymbol{\ell}_{t,k}\in\Rb^{\Ntx B_s}.
    \label{eq:soft_receiver}
\end{equation}
For a coded bit $b_{t,k,s,q}$ at stream index $s$ and bit position $q$, the corresponding posterior LLR is
\begin{equation}
    \ell^{\star}_{t,k,s,q}
    =\log\frac{\Pr(b_{t,k,s,q}=1\mid\mathcal{Y},\mathcal{C}_{\mathrm p})}
                   {\Pr(b_{t,k,s,q}=0\mid\mathcal{Y},\mathcal{C}_{\mathrm p})}.
    \label{eq:llr_target}
\end{equation}
Positive LLRs favor bit one, and the soft outputs are passed to the channel decoder.
Pilot-derived channel estimates can be included as deterministic features of $\mathcal{C}_{\mathrm p}$ without requiring the true channel at inference.

In the ICL interpretation, known pilot symbols paired with their received observations provide examples of the current channel mapping, while data observations form the queries~\cite{zecchin2023mimoicl,zecchin2025context}.
The parameters $\theta$ are learned offline; the current context conditions the forward pass without an online parameter update.
The bandwidth-reuse objective is to apply the same $\theta$ when $\Nsc$ changes, with the context and output sizes following the new allocation.
Section~\ref{ssec:token} specifies how these observations and context roles are represented by tokens.

\subsection{Self-Attention and Axial Factorization}
\label{ssec:prelim_axial}

Let $\Zv\in\Rb^{N\times d}$ contain $N$ tokens of feature width $d$.
For one attention head, learned projections produce $\Qv=\Zv\Wv_Q$, $\Kv=\Zv\Wv_K$, and $\Vv=\Zv\Wv_V$, with head dimension $d_h$.
Scaled dot-product attention computes
\begin{equation}
    \operatorname{Attn}(\Zv)
    =\softmax\!\left(\frac{\Qv\Kv^{\top}}{\sqrt{d_h}}\right)\Vv,
    \label{eq:self_attention}
\end{equation}
where softmax is applied row-wise.
Multi-head self-attention concatenates the head outputs and applies an output projection; Transformer layers also use position-wise feed-forward processing, residual connections, and normalization~\cite{vaswani2017attention}.
Positional information distinguishes token locations, for example through additive embeddings or rotary transformations of queries and keys~\cite{su2024roformer}.

For a grid of $\Nsym\times\Nsc$ tokens, full attention treats $N=\Nsym\Nsc$ positions as one sequence.
Its pairwise interaction cost is $\mathcal{O}(\Nsym^2\Nsc^2d)$.
Axial attention processes one grid axis at a time~\cite{ho2019axial}: time-axis attention operates separately at each subcarrier, and frequency-axis attention operates separately at each OFDM symbol.
A pair of these operations has interaction cost
\begin{equation}
    \mathcal{O}\!\left(\Nsc\Nsym^2d+\Nsym\Nsc^2d\right).
    \label{eq:axial_cost}
\end{equation}
These expressions count the query-key and attention-value interactions; projections and feed-forward layers contribute additional computation that is linear in the number of REs for fixed feature widths.
Although axial factorization reduces the cost of full-grid attention, the cost of unrestricted frequency attention still grows quadratically with $\Nsc$.
This remaining dependence motivates the bounded frequency interactions developed in Section~\ref{ssec:axial}.

\section{Method: \ourmethods{}}
\label{sec:method}

In this section, we present the architecture and training procedure of \ourmethods{}. The receiver represents received observations and pilot context as RE tokens and maps them to bit LLRs through alternating time-axis and local frequency-axis attention. We also describe how the architecture supports bandwidth reuse and an optional CE+SD extension.

\subsection{Per-RE Input Representation and Pilot Context}
\label{ssec:token}

\begin{figure}[t]
    \centering
    \includegraphics[width=0.85\columnwidth]{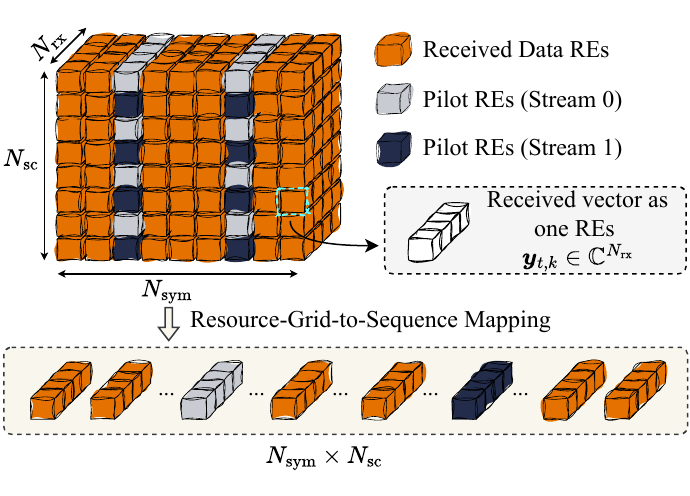}
    \caption{Resource-grid-to-sequence mapping. Each of the $\Nsym\Nsc$ sequence elements contains the complete receive vector $\yv_{t,k}\in\Cb^{\Nrx}$ at one RE. Colors distinguish data REs and pilot REs assigned to the two streams. The mapping preserves the time-frequency coordinates of each RE. Grid dimensions are schematic.}
    \label{fig:method_grid}
\end{figure}

Each RE is represented by one token containing features of its complete receive vector (Fig.~\ref{fig:method_grid}). Pilot tokens additionally encode the known transmitted symbols and pilot-derived channel features.
Let $\rho(\cdot)$ denote a fixed real-valued vectorization that stacks the real and imaginary components of a complex vector or matrix.
The received feature $\rho(\yv_{t,k})\in\Rb^{2\Nrx}$ is available at every RE.
At pilot REs, the known transmitted vector $\rho(\xv_{t,k})\in\Rb^{2\Ntx}$ provides the corresponding input to the channel mapping.
Its zero entries encode the pilot allocation information.
Data symbols are unknown and are not supplied to the tokenizer.

We additionally provide pilot-derived LS features~\cite{coleri2002channel} as a deterministic summary of the context.
At an RE assigned to pilot stream $s$, the observed channel column is estimated as
\begin{equation}
    \widehat{\mathbf h}^{\mathrm{LS},(s)}_{t,k}
    =\frac{\yv_{t,k}(x^{(s)}_{t,k})^*}{|x^{(s)}_{t,k}|^2}.
    \label{eq:method_pilot_ls}
\end{equation}
The implementation constructs a complete matrix $\hls_{t,k}\in\Cb^{\Nrx\times\Ntx}$ using nearest-neighbor filling from each stream's pilot observations, as in the preprocessing of DeepRx-MIMO~\cite{korpi2021deeprx}.
Only the matrices at pilot REs enter the tokenizer.
Consequently, a pilot token contains all channel-column features, including filled entries, rather than only the column observed directly at that RE.
These features do not require the true channel.

Let $m_{t,k}$ equal one at pilot REs and zero at data REs.
With learned affine projections $f_y$, $f_x$, and $f_h$ into $\Rb^d$, the initial token is
\begin{equation}
\begin{aligned}
    \zv^{(0)}_{t,k}
    ={}&f_y\big(\rho(\yv_{t,k})\big)
       +\mathbf e^{\mathrm{role}}_{m_{t,k}}+\mathbf e^{\mathrm{time}}_t\\
      &+m_{t,k}\!\left[f_x\big(\rho(\xv_{t,k})\big)
             +f_h\big(\rho(\hls_{t,k})\big)\right],
\end{aligned}
    \label{eq:method_token}
\end{equation}
where the bracketed terms are evaluated only at pilot REs.
The role embedding distinguishes known context from unknown data, and the time embedding identifies the OFDM-symbol position.
No learned absolute frequency embedding is added.
The projection dimensions depend on the antenna configuration and feature width, but not on the number of allocated subcarriers.

The $\Nre=\Nsym\Nsc$ tokens form a tensor of size $\Nsym\times\Nsc\times d$, with time and frequency coordinates preserved when flattened for storage.

\subsection{Local Axial Processing and Soft Detection}
\label{ssec:axial}

\begin{figure*}[t]
    \centering
    \includegraphics[width=\textwidth]{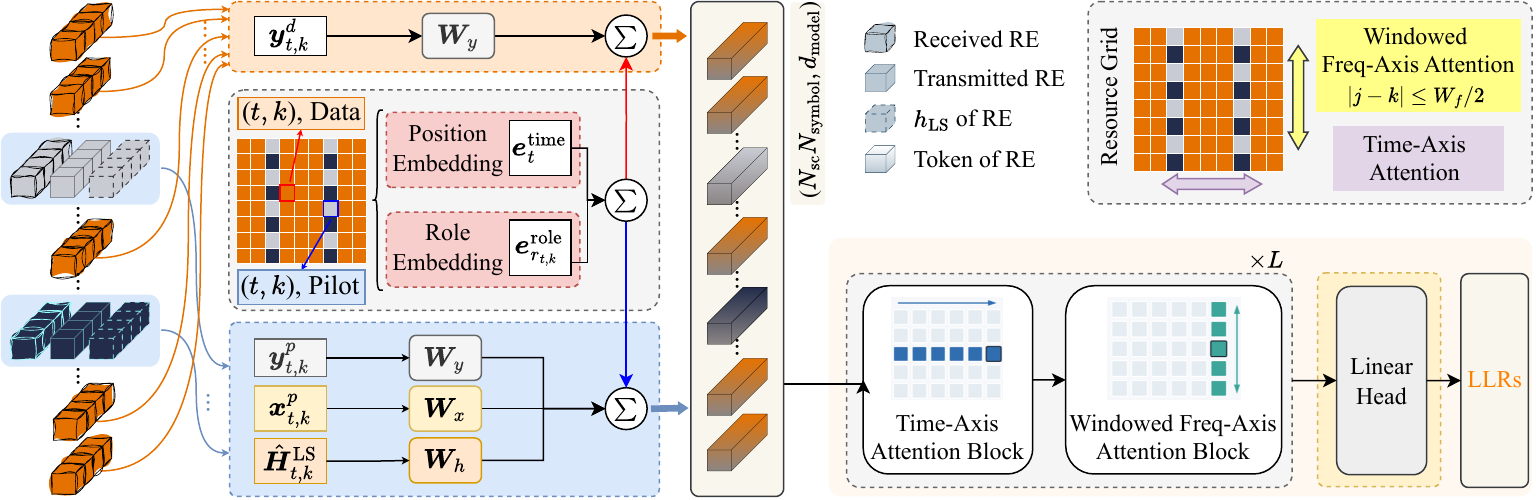}
    \caption{\ourmethods{} architecture. RE tokens combine received observations, pilot context, and role and time embeddings. Five pairs of time-axis and windowed frequency-axis attention layers process the grid, followed by the LLR head.}
    \label{fig:method_architecture}
\end{figure*}

The distinct time and frequency axes of the OFDM grid guide the interaction structure: the detector alternates time-axis and frequency-axis attention~\cite{ho2019axial}, preserving the grid organization introduced by the tokenizer (Fig.~\ref{fig:method_architecture}).
A time layer processes all $\Nsym$ symbols separately at each subcarrier, allowing data observations to interact with the pilot-bearing symbols.
A frequency layer uses sliding-window attention~\cite{beltagy2020longformer} over neighboring subcarriers separately at each OFDM symbol.
For frequency radius $r_f$, define
\begin{equation}
    \mathcal{W}(k)=\{j:0\leq j<\Nsc,\ |j-k|\leq r_f\},
    \qquad \Wf=2r_f+1.
    \label{eq:method_window}
\end{equation}
At fixed subcarrier spacing $\Delta f$, the radius $r_f$ corresponds to a maximum frequency offset of $r_f\Delta f$ from the query, independent of the total PRB allocation. The window therefore defines a consistent physical neighborhood across bandwidths while bounding the number of interactions per query.
At allocation boundaries, only valid neighbors participate in the softmax.
Successive frequency layers expand the effective receptive field, so the final prediction can depend on observations beyond a single layer's window.

Rotary positional encoding (RoPE) is applied to queries and keys along the active axis~\cite{su2024roformer}.\footnote{An auxiliary ablation at 4 and 16 PRBs, with time-axis RoPE and local attention unchanged, showed modest pre-decoding BER improvements from frequency-axis RoPE. Post-decoding BER favored its removal at high SNR.}
For one frequency-attention head, let $\mathbf q_{t,k}$, $\boldsymbol\kappa_{t,k}$, and $\mathbf v_{t,k}$ denote the projected query, key, and value, respectively.
Writing the rotary transformation at index $k$ as $\mathbf R(k)$, scaled dot-product attention~\cite{vaswani2017attention} gives
\begin{equation}
\begin{aligned}
    a_{t,k,j}
    & =\frac{\exp\!\left(\mathbf q_{t,k}^{\top}\mathbf R(j-k)
                          \boldsymbol\kappa_{t,j}/\sqrt{d_h}\right)}
                 {\displaystyle\sum_{u\in\mathcal W(k)}
                   \exp\!\left(\mathbf q_{t,k}^{\top}\mathbf R(u-k)
                          \boldsymbol\kappa_{t,u}/\sqrt{d_h}\right)},\\
    \mathbf o_{t,k}
    & =\sum_{j\in\mathcal W(k)}a_{t,k,j}\mathbf v_{t,j}.
\end{aligned}
    \label{eq:method_local_attention}
\end{equation}
Here $\mathbf R(k)^\top\mathbf R(j)=\mathbf R(j-k)$ makes the positional contribution depend on the relative displacement.
Time attention uses the analogous rotary transformation over the full symbol axis.
Head outputs are concatenated and projected back to width $d$.
The frequency implementation gathers local keys and values without constructing a full $\Nsc\times\Nsc$ attention matrix.

Each axial layer uses layer normalization~\cite{ba2016layer} in the pre-normalization arrangement~\cite{xiong2020layer}, together with residual connections~\cite{he2016deep}:
\begin{equation}
\begin{aligned}
    \mathbf U^{(\ell)}
      &=\mathbf Z^{(\ell-1)}
        +\operatorname{MHA}_{a_\ell}\!\left(
          \operatorname{LN}_{\ell,1}(\mathbf Z^{(\ell-1)})\right),\\
    \mathbf Z^{(\ell)}
      &=\mathbf U^{(\ell)}
        +\operatorname{FFN}_{\ell}\!\left(
          \operatorname{LN}_{\ell,2}(\mathbf U^{(\ell)})\right),
\end{aligned}
    \label{eq:method_layer}
\end{equation}
where $a_\ell$ identifies the active axis.
The position-wise feed-forward network comprises affine maps from $d$ to $d_{\mathrm{ff}}$ and back, with a Gaussian error linear unit (GELU) activation~\cite{hendrycks2016gelu} between them.
Layer parameters are shared across grid positions, with separate parameters for different layers.

\ourmethods{} starts with a time layer and contains five time layers and five frequency layers.
We use $d=64$, four heads, $d_{\mathrm{ff}}=256$, and $r_f=12$, giving up to 25 attended subcarriers per frequency query.
After the final layer, a layer normalization and a shared two-layer GELU head map each data token from $d$ to $d$ and then to $\Ntx B_s$ bit LLRs:
\begin{equation}
    \boldsymbol\ell_{t,k}
      =g_{\mathrm{LLR}}\!\left(\operatorname{LN}_{\mathrm{out}}
                    (\zv^{(\nsd)}_{t,k})\right),
    \qquad (t,k)\in\mathcal G_{\mathrm d}.
    \label{eq:method_llr}
\end{equation}
The head has no output sigmoid, so its logits are supplied directly as soft information to the decoder.
The primary architecture has no intermediate channel-estimation head or explicit equalization stage.

\subsection{Bandwidth Reuse and Computational Complexity}
\label{ssec:method_complexity}

The architecture separates the number of processed REs from the dimensions of its learned mappings.
When $\Nsc$ changes, the receiver rebuilds the pilot and data masks, token coordinates, and output indexing for the new grid.
The input projections, role and time embeddings, axial layers, and output head use the same weights at every bandwidth.
Rotary transformations are generated at the required coordinates and require no learned frequency table tied to the training allocation.

The bounded neighborhood in Section~\ref{ssec:axial} controls computation without assuming that channel correlation vanishes outside the window.
With fixed depth, distant frequency observations outside the accumulated receptive field do not directly interact, and the window size controls the resulting performance-computation tradeoff.
These structural properties permit execution with unchanged weights across bandwidths; the accuracy achieved after such reuse is assessed experimentally.

Let $L_t$ and $L_f$ denote the numbers of time and frequency layers, with $L=L_t+L_f$.
Including the attention interactions, linear projections, and feed-forward networks, the backbone cost scales as
\begin{equation}
\begin{aligned}
    \mathcal C_{\mathrm{backbone}}=\mathcal O\big(&
       L_t\Nsc\Nsym^2d
       +L_f\Nsym\Nsc\Wf d\\
       &+L\Nsym\Nsc(d^2+d\,d_{\mathrm{ff}})\big).
\end{aligned}
    \label{eq:method_complexity}
\end{equation}
Token projections and the output head add computation linear in the number of REs for fixed input and output dimensions.
Thus, with fixed symbol count, widths, depth, and frequency radius, the neural receiver cost grows linearly with $\Nsc$, and its parameter count is independent of $\Nsc$.
This removes the quadratic frequency interaction term in \eqref{eq:axial_cost}.
It does not by itself establish a lower operation count than another receiver whose cost also scales linearly.
The experimental comparison therefore counts the operations of the complete evaluated neural models under a common convention, and reports runtime separately from this architectural scaling analysis.

\subsection{Optional CE+SD Extension}
\label{ssec:arch}

\begin{figure}[!tb]
    \centering
    \includegraphics[width=\columnwidth]{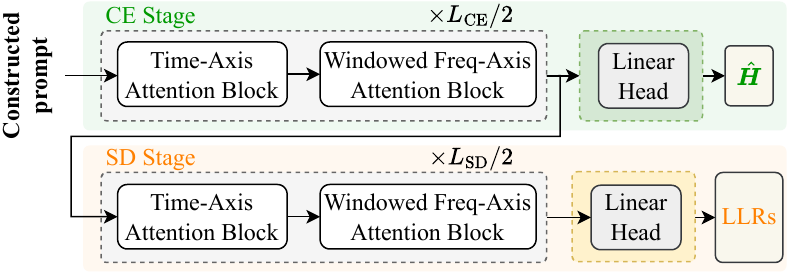}
    \caption{Optional CE+SD extension. Intermediate CE features feed both the channel-estimation head and the SD stage, which produces bit LLRs.}
    \label{fig:method_ce_sd}
\end{figure}

An explicit channel estimate can be added by partitioning the axial stack into a CE stage and an SD stage, as illustrated in Fig.~\ref{fig:method_ce_sd}.
The CE stage transforms the same input tokens through $\nce$ axial layers, and its normalized output $\mathbf Z_{\mathrm{ce}}$ feeds both a channel head and the subsequent detector.
The channel head applies a two-layer GELU mapping from $d$ to $d$ and then to $2\Nrx\Ntx$ at every RE, which is reshaped into $\hest_{t,k}$.
The detector applies a further $\nsd$ axial layers to $\mathbf Z_{\mathrm{ce}}$, followed by the LLR head at data REs.

The detector receives the intermediate features rather than an explicitly equalized signal constructed from $\hest$.
Accordingly, channel estimation serves as an auxiliary supervised task~\cite{liebel2018auxiliary}, without constraining the detector to a prescribed equalization rule.
Detection gradients propagate through the CE stage, with no gradient stop between stages.
The evaluated extension uses four CE layers and six SD layers, for a total of ten axial layers.
Its two CE frequency layers use radii 12 and 6, while its three SD frequency layers use radius 12.
Section~\ref{ssec:ce_extension} evaluates the channel estimates and compares detection BER with a detection-only variant trained under the same budget.

\subsection{Training Objective and Inference}
\label{ssec:loss}

The receiver is trained offline using independently and uniformly sampled modulation bits.
For training example $n$, let $K$ be the number of data bits and let $b_{n,i}$ and $\ell_{n,i}$ denote a bit label and its predicted logit.
The mean binary cross-entropy~\cite{goodfellow2016deep} is
\begin{equation}
    \overline{\mathcal L}_n
      =\frac{1}{K}\sum_{i=1}^{K}
        \left[\log(1+\exp(\ell_{n,i}))-b_{n,i}\ell_{n,i}\right].
    \label{eq:method_bce}
\end{equation}
Following the SNR weighting used in DeepRx-MIMO~\cite{korpi2021deeprx}, we assign greater weight to examples at higher receive SNR and normalize by the sum of weights:
\begin{equation}
    \Ldet=\frac{\sum_n w_n\overline{\mathcal L}_n}{\sum_n w_n},
    \qquad w_n=\log_2(1+10^{\gamma_n/10}),
    \label{eq:method_weighted_bce}
\end{equation}
where $\gamma_n$ is the SNR in decibels.
The detection-only model minimizes $\Ldet$ and does not require channel labels.

For the optional CE+SD extension, simulated channel labels additionally supervise the real and imaginary components of the channel estimate.
With batch size $B$, the auxiliary loss is the mean-squared error~\cite{goodfellow2016deep} over these components:
\begin{equation}
    \Lceaux=\frac{1}{2B\Nre\Nrx\Ntx}
       \sum_{n=1}^{B}\sum_{(t,k)\in\mathcal G}
       \left\|\hest^{(n)}_{t,k}-\Hv^{(n)}_{t,k}\right\|_F^2.
    \label{eq:method_ce_loss}
\end{equation}
Unlike the detection loss, this mean-squared error has no SNR weighting.
The joint objective is $\Ltotal=\Ldet+\lambda_e\Lceaux$.
We adapt the cosine-annealing form in~\cite{loshchilov2017sgdr} to decrease the auxiliary-loss weight over training:
\begin{equation}
    \lambda_e=\lambda_{\min}
      +\frac{\lambda_0-\lambda_{\min}}{2}
        \left[1+\cos\!\left(\frac{\pi e}{E-1}\right)\right],
    \label{eq:method_ce_schedule}
\end{equation}
for epoch $e\in\{0,\ldots,E-1\}$.
We set $\lambda_0=1$ and $\lambda_{\min}=0.05$.
Both objectives are active from the first epoch; there is no CE-only training stage.

Training uses a narrow allocation and samples the pilot-symbol positions according to the protocol in Section~\ref{ssec:setup}.
This exposes the receiver to varying context locations without changing the meaning of pilot and data tokens.
At inference, the weights are frozen and the receiver constructs its context from the current received pilots and their known transmitted symbols.
It then produces LLRs in one forward pass, without online parameter optimization.
Hard decisions are used for pre-decoding BER, and the same soft outputs are passed to the channel decoder for post-decoding BER.
The optional channel head supplies $\hest$ when an explicit channel output is requested.
\space 

\section{Experimental Results}
\label{sec:results}
\subsection{Experimental Setup and Evaluation Protocol}
\label{ssec:setup}

We evaluate reception accuracy, computational cost, and frozen-weight reuse beyond the 4-PRB UMa training configuration. Window-range and NRX-depth controls examine the effect of local processing and the comparison with lower-cost baseline variants.
Unless otherwise specified, the experiments use the single-user uplink configuration in Table~\ref{tab:setup}.
The link simulation uses Sionna~\cite{hoydis2022sionna}, and the neural receivers are implemented in PyTorch.
Channel topologies are generated independently according to 3GPP TR~38.901~\cite{3gpp38901}.
We use normalized frequency-domain channels with variation across OFDM symbols and ideal synchronization; path loss, shadow fading, intercell interference, and intercarrier and intersymbol interference are disabled in this reference setting.

\begin{table}[t]
    \centering
    \caption{Default System and Training Configuration}
    \label{tab:setup}
    \small
    \setlength{\tabcolsep}{4pt}
    \renewcommand{\arraystretch}{1.12}
    \begin{tabular}{@{}p{0.46\columnwidth}p{0.48\columnwidth}@{}}
        \toprule
        Parameter & Setting \\
        \midrule
        Transmit / receive antennas & 2 / 4 \\
        Spatial streams / modulation & 2 / 64-QAM \\
        Carrier / subcarrier spacing & 2.6~GHz / 30~kHz \\
        Training PRBs / subcarriers & 4 / 48 effective \\
        OFDM symbols / pilot symbols & 14 / 2 \\
        Training FFT size / CP length & 64 / 6 samples \\
        Training channel / speed & Urban macrocell (UMa) / $\mathcal{U}(0,35)$~m/s \\
        Training receive SNR & $\mathcal{U}(0,45)$~dB \\
        Attention layers / heads & 10 alternating layers / 4 \\
        Model / feed-forward width & 64 / 256 \\
        Frequency-attention radius & 12 subcarriers \\
        Training epochs / updates & 500 / 100,000 \\
        Effective training batch size & 112 \\
        Target LDPC code rate & $517/1024$ \\
        LDPC decoding iterations & 20 \\
        \bottomrule
    \end{tabular}
\end{table}

The spatial streams use frequency-orthogonal pilots on the designated pilot-bearing symbols.
During training, their zero-based symbol indices are sampled uniformly from six pairs: $\{ (2,11), (3,10), (4,9), (5,8), (1,12), (0,13) \}$.
Evaluation uses $(2,11)$, $(3,10)$, $(5,8)$, and $(0,13)$, all within the training set of layouts.
The SNR is the expected total desired signal power per receive antenna, averaged over data resource elements, divided by the noise power.
Training SNR and user speed are sampled independently.

The main comparisons use the detection-only \ourmethods{} architecture described in Section~\ref{ssec:axial}.
We compare against NRX-unfold, our implementation of the CNN/GNN receiver in~\cite{cammerer2023neural} with three message-passing steps and separate parameters across steps, DeepRx-MIMO~\cite{korpi2021deeprx}, and EqDeepRx~\cite{honkala2026eqdeeprx}.
We also include two training-free references: LS-LMMSE, which applies least-squares channel estimation with linear interpolation followed by linear minimum mean-square error equalization, and perfect-CSI LMMSE, which uses the true channel with zero channel-estimation error variance.
Both use a posteriori probability 64-QAM demapping and the same LDPC decoder as the neural receivers.

All neural receivers are trained from scratch under the same data distribution and sample budget, using independent random bits, the SNR-weighted detection objective in \eqref{eq:method_weighted_bce}, and model-specific optimizer settings.
EqDeepRx additionally uses its symbol-estimation and activation-regularization losses.
\ourmethods{}, NRX-unfold, and DeepRx-MIMO use AdamW with initial learning rates $8\times10^{-4}$, $10^{-3}$, and $5\times10^{-4}$, respectively, and 500 warmup updates.
EqDeepRx uses LAMB with the initial learning rate of $4.4\times10^{-3}$ reported in~\cite{honkala2026eqdeeprx}; we use no warmup for this receiver.
The learning rate then decays linearly to zero for each model.
The reported main comparisons use one fixed training run per model (seed~0), with paired training-data seeds and the checkpoint from the final training epoch.
Bandwidth and channel-transfer tests reuse these weights without fine-tuning or target-domain checkpoint selection.

Independent evaluation uses 5G low-density parity-check (LDPC) coded transmissions.
At the training bandwidth, each stream carries 1744 information bits in a 3456-bit codeword; in the single-codeword 16-PRB evaluation, the corresponding lengths are 6979 and 13,824 bits.
All receivers use the same flooding decoder with boxplus-phi check-node updates and log-likelihood ratios clipped to $[-20,20]$.
Reception performance is evaluated using pre-decoding bit error rate (BER) and post-decoding BER.
Pre-decoding BER compares hard decisions on the receiver output log-likelihood ratios with the transmitted coded bits, whereas post-decoding BER compares the decoded information bits with the transmitted information bits.
Both metrics are computed from the same coded transmissions.
All receivers are evaluated on paired transmissions with the same data and pilot configurations, using adaptive sampling of 3,200--51,200 realizations per test condition.
Reported BER curves average the error rates equally over the four pilot layouts.
Subsequent experiments specify departures from this default configuration, first isolating bandwidth changes under UMa and then evaluating unseen channels at the training and extended bandwidths.

\begin{figure}[!htb]
    \centering
    \includegraphics[width=\columnwidth]{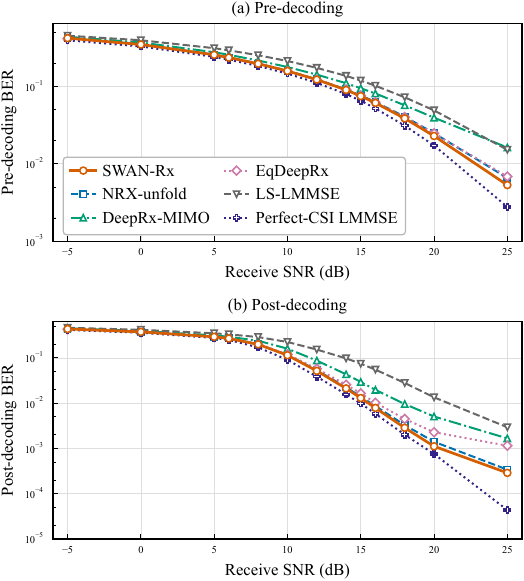}
    \caption{Reception on UMa at 4 PRBs: (a) pre-decoding BER; (b) post-decoding BER.}
    \label{fig:reception_performance}
\end{figure}
 
\subsection{Reception Performance and Generalization}
\label{ssec:performance_generalization}
\subsubsection{Lower BER at the Training Bandwidth}
\label{ssec:ber_snr}
\label{sssec:ber_indomain}
We first evaluate independent coded UMa transmissions at 4 PRBs. All receivers use a common SNR grid, with additional points between 6 and 18~dB to resolve the decoding transition.

The results in Fig.~\ref{fig:reception_performance} show lower pre- and post-decoding BER for \ourmethods{} than for the implemented three-step NRX and DeepRx-MIMO receivers over the plotted SNR grid.
EqDeepRx remains competitive at low SNR, whereas the post-decoding curves separate in favor of \ourmethods{} as SNR increases.
Compared with LS-LMMSE, \ourmethods{} achieves lower pre- and post-decoding BER at every plotted SNR, while perfect-CSI LMMSE indicates the remaining gap to the linear receiver supplied with the true channel.

\subsubsection{Reception up to 273 PRBs Without Retraining}
\label{sssec:ber_prb_extrap}
To test reception beyond the training bandwidth, we increase the allocation to 4, 8, 16, 64, 128, and 273 PRBs while keeping the UMa channel model and all learned weights fixed. The largest allocation is 68.25 times the training bandwidth. Scaling the FFT size and CP length together holds the OFDM-symbol duration constant.

All allocations follow a common LDPC packing rule: each stream is filled with independent $(k,n)=(3489,6912)$ codeblocks, followed by a shorter block if required to fill the remaining data positions.
For a shorter block of $n_{\mathrm{tail}}$ coded bits, the information length is $\lfloor n_{\mathrm{tail}}(517/1024)\rfloor$.
Thus, four PRBs use one $(1744,3456)$ block, while 8/16/64/128 PRBs use 1/2/8/16 full blocks per stream, respectively.
At 273 PRBs, 34 full blocks and one $(436,864)$ block occupy all data positions.
All blocks use the same decoder configuration, and all transmitted bits contribute to the BER metrics.
This mapping uses independent codeblocks without transport-block CRC or NR transport-block segmentation; post-decoding BER also reflects the shorter blocks where present.

\begin{figure}[!htb]
    \centering
    \includegraphics[width=\columnwidth]{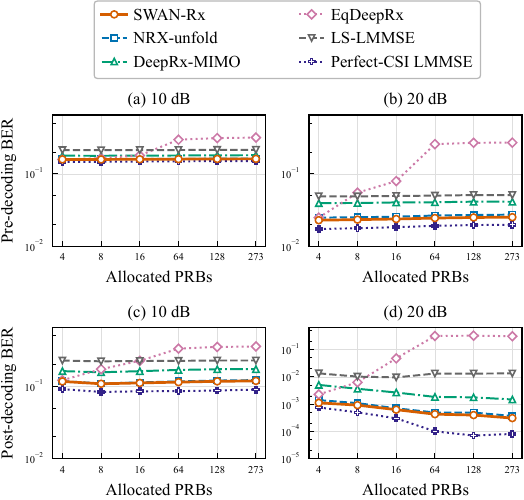}
    \caption{Frozen-weight bandwidth extrapolation from 4 PRBs on UMa. Columns: 10 and 20~dB; (a)/(b): pre-decoding BER; (c)/(d): post-decoding BER. The dotted line marks the training bandwidth.}
    \label{fig:ber_prb_extrap}
\end{figure}
 Across all six allocations in Fig.~\ref{fig:ber_prb_extrap}, \ourmethods{} achieves lower pre- and post-decoding BER than NRX and DeepRx-MIMO at both plotted SNRs.
At 273 PRBs and 20~dB, its post-decoding BER is $3.07\times10^{-4}$, compared with $3.68\times10^{-4}$ for NRX.

These results show that the allocation-independent processing in Section~\ref{ssec:method_complexity} supports bandwidth reuse while preserving the measured BER advantage over NRX and DeepRx-MIMO. The window-range control in Fig.~\ref{fig:windowing} examines how the frequency-interaction range affects this transfer.

\begin{figure}[!htb]
    \centering
    \includegraphics[width=\columnwidth]{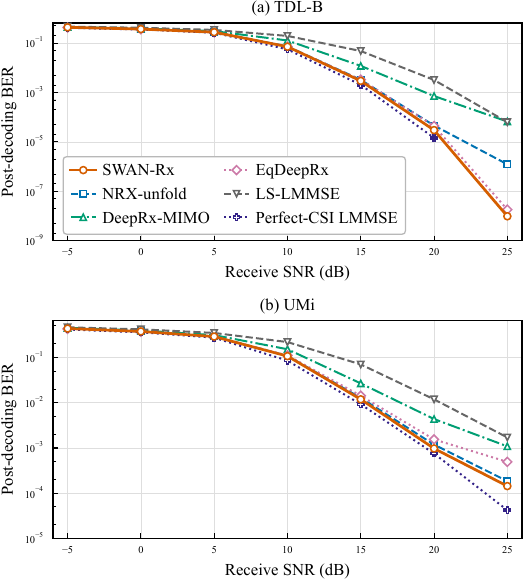}
    \caption{Post-decoding BER under frozen-weight transfer from UMa to (a) TDL-B and (b) UMi at 4 PRBs.}
    \label{fig:ood_statistical_4prb}
\end{figure}
 \subsubsection{Transfer to Unseen Channels Without Retraining}
\label{sssec:ber_zeroshot}
With the 4-PRB UMa-trained weights frozen, we test channel transfer to TDL-B and UMi at 4 PRBs, followed by joint channel and bandwidth transfer to CDL-B, TDL-B, UMi, NewYork-NLoS, and I1 at 16 PRBs.
CDL-B and TDL-B use a speed of 10~m/s and a delay spread of 100~ns; UMi uses the speed distribution in Table~\ref{tab:setup}.
Each ray-tracing scenario replays 256 fixed users with channels that are static within a slot.

\begin{figure*}[!tbp]
    \centering
    \includegraphics[width=\textwidth]{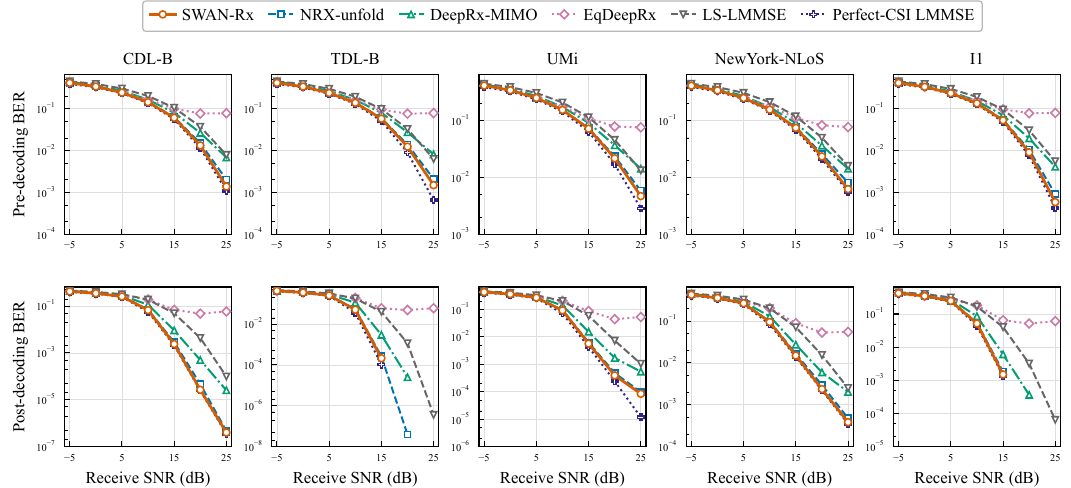}
    \caption{Frozen-weight transfer from 4-PRB UMa training to five unseen channels at 16 PRBs. Top: pre-decoding BER; bottom: post-decoding BER.}
    \label{fig:ood_16prb_combined}
\end{figure*}
 At four PRBs, \ourmethods{} achieves lower post-decoding BER than NRX and DeepRx-MIMO on both TDL-B and UMi over the plotted SNR grid (Fig.~\ref{fig:ood_statistical_4prb}).
Across the five channels in Fig.~\ref{fig:ood_16prb_combined}, \ourmethods{} achieves lower pre-decoding BER than NRX and DeepRx-MIMO over the evaluated SNR grid.
Its post-decoding BER is also no higher than either receiver, with ties where no errors are observed.
For example, at 20~dB on NewYork-NLoS, its post-decoding BER is $2.37\times10^{-3}$, compared with $2.96\times10^{-3}$ for NRX and $5.96\times10^{-3}$ for DeepRx-MIMO.
The reception advantage thus extends to joint changes in bandwidth and channel conditions within the tested scenarios and user banks, without additional training.
In comparison, EqDeepRx exhibits pronounced high-SNR error floors in several of these channel-transfer scenarios, consistent with its degraded performance at larger allocations in Fig.~\ref{fig:ber_prb_extrap}.
Points with zero observed errors are omitted from the logarithmic curves and do not establish zero error probability.

\begin{figure}[!htb]
    \centering
    \includegraphics[width=\columnwidth]{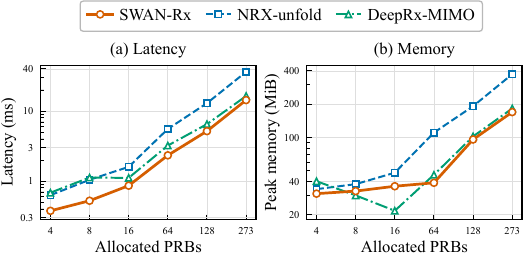}
    \caption{Optimized inference at batch size 1 over 4--273 PRBs: latency (left) and peak allocated memory (right).}
    \label{fig:optimized_efficiency}
\end{figure}
 \subsection{Computational Complexity and Inference Efficiency}
\label{ssec:deployment}
\subsubsection{Lower Forward-Pass Computation}
\label{sssec:forward_cost}
Table~\ref{tab:computational_cost} reports parameter counts and forward-pass operations for the receiver checkpoints evaluated in Fig.~\ref{fig:reception_performance}. At 4 PRBs, \ourmethods{} requires 0.726 GFLOPs per sample, 56.3\% less than three-step NRX and 35.7\% less than DeepRx-MIMO, with fewer parameters than either baseline. These reductions accompany lower pre- and post-decoding BER under the common training protocol.

Increasing the allocation from four to sixteen PRBs raises its forward-pass cost to 2.906 GFLOPs per sample, approximately fourfold, while its 507,916 parameters remain unchanged. This agrees with the linear bandwidth scaling derived in Section~\ref{ssec:method_complexity}.
The counts include attention, normalization, and activation operations; the arithmetic convention is specified in the supplementary material.
One- and two-step NRX use fewer parameters, and Fig.~\ref{fig:nrx_budget} compares their reception performance with that of \ourmethods{}.
\begin{table}[!t]
\centering
\caption{Parameter counts and forward-pass computation. NRX depth variants are separately trained.}
\label{tab:computational_cost}
\footnotesize
\setlength{\tabcolsep}{4pt}
\begin{tabular}{lrrr}
\toprule
Receiver & Parameters & \multicolumn{2}{c}{GFLOPs/sample} \\
\cmidrule(lr){3-4}
 & & 4 PRBs & 16 PRBs \\
\midrule
\ourmethods{} & 507,916 & 0.726 & 2.906 \\
NRX, 1 step & 270,342 & 0.737 & 2.949 \\
NRX, 2 steps & 440,838 & 1.200 & 4.801 \\
NRX, 3 steps & 611,334 & 1.663 & 6.654 \\
DeepRx-MIMO & 842,588 & 1.130 & 4.519 \\
\bottomrule
\end{tabular}
\end{table}
 
\subsubsection{Faster Inference After Implementation Optimization}
\label{sssec:runtime}
We then measure execution time and peak allocated memory over 4--273 PRBs to assess the cost of the implemented receivers. For each bandwidth and batch size, all receivers are measured within the same GPU allocation, from inputs that already include LS estimates to output LLRs.
Measurements use an NVIDIA RTX PRO 6000 Blackwell MIG 1g.24gb instance with FP32/complex64 arithmetic and TF32 disabled. Peak allocated memory includes CUDA graph workspace where used.
Channel generation, external LS estimation, and LDPC decoding are excluded from both runtime measurements and FLOP counts.
Each receiver uses the fastest stable implementation found in a search over model-specific kernels, compilation, and CUDA graph execution; batch-32 candidates also include processing in smaller subbatches.
Latency and memory are reported for the same implementation, with the search settings and timing procedure given in the supplementary material.

At batch size 1, \ourmethods{} has 40.2--60.5\% lower measured latency than NRX and 12.4--53.3\% lower latency than DeepRx-MIMO across the six bandwidths (Fig.~\ref{fig:optimized_efficiency}).
Its peak allocation is lower than that of NRX throughout, while DeepRx-MIMO uses less memory at 8 and 16 PRBs.
The memory curves also reflect changes in execution mode and temporary allocations: DeepRx-MIMO uses CUDA graphs at 4 PRBs but not at 8 or 16 PRBs, and \ourmethods{} uses graphs at 16 and 128 PRBs but not at 64 PRBs.
Table~\ref{tab:runtime_batch32} shows higher throughput and lower peak allocation for \ourmethods{} at both batch-32 bandwidth endpoints.
\begin{table}[!t]
\centering
\caption{Throughput and peak allocated memory at batch size 32. $\dagger$ indicates more decoded bit errors than the original implementation in at least one validation condition.}
\label{tab:runtime_batch32}
\footnotesize
\setlength{\tabcolsep}{3pt}
\begin{tabular}{rlrr}
\toprule
PRBs & Receiver & Samples/s & Memory (MiB) \\
\midrule
4 & \ourmethods{}$^{\dagger}$ & 6028.6 & 59.9 \\
4 & NRX, 3 steps & 3076.3 & 79.3 \\
4 & DeepRx-MIMO$^{\dagger}$ & 4819.8 & 82.5 \\
273 & \ourmethods{}$^{\dagger}$ & 61.5 & 1165.4 \\
273 & NRX, 3 steps & 27.3 & 10777.7 \\
273 & DeepRx-MIMO$^{\dagger}$ & 56.2 & 4730.8 \\
\bottomrule
\end{tabular}
\end{table}
 
The measured runtime advantage depends on these optimizations.
In separate measurements of the original PyTorch implementations without compilation at 4 and 16 PRBs, \ourmethods{} has higher latency and peak allocation than NRX and DeepRx-MIMO at both batch sizes.
After implementation selection, paired validation used 1536 coded samples per configuration, covering four pilot layouts, three channels, and four SNRs. For \ourmethods{} at batch size 1, these checks found no newly failed block and no per-condition increase in decoded bit errors at any of the six bandwidths in Fig.~\ref{fig:optimized_efficiency}. Some optimized baseline and batch-32 configurations showed increases in decoded bit errors. These differences are reported in the supplementary material and identified for the batch-32 results in Table~\ref{tab:runtime_batch32}; decoding performance is therefore not assumed to be unchanged by optimization.

\begin{figure}[!htb]
    \centering
    \includegraphics[width=\columnwidth]{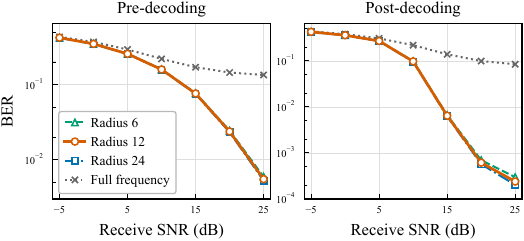}
    \caption{Frequency-attention range comparison on UMa at 16 PRBs with weights trained at 4 PRBs. Left: pre-decoding BER; right: post-decoding BER.}
    \label{fig:windowing}
\end{figure}
 \Needspace{6\baselineskip}
\subsection{Architectural Ablations}
\label{ssec:ablations}
The following controls examine the effect of frequency-attention range on bandwidth transfer and whether reducing NRX depth changes the reception--computation comparison.
\subsubsection{Local Attention Improves Bandwidth Transfer}
The frequency window determines which observations can interact within a layer. To examine its effect on bandwidth transfer, we train parameter-matched models with radii 6, 12, and 24 and with full-frequency attention, using the same four-PRB training allocation and update budget. Radii 6, 12, and 24 allow at most 13, 25, and 49 subcarriers per query, respectively; full-frequency attention uses all allocated subcarriers. We then evaluate the frozen models at sixteen PRBs on UMa.

The local variants achieve substantially lower BER than full-frequency attention at the unseen allocation (Fig.~\ref{fig:windowing}). At 20~dB, the full-frequency model's pre-decoding BER rises from $2.21\times10^{-2}$ at four PRBs to $1.44\times10^{-1}$ at sixteen PRBs, while the radius-12 model changes from $2.29\times10^{-2}$ to $2.37\times10^{-2}$. This contrast supports local frequency processing for bandwidth transfer.
Radius 24 gives slightly lower BER than radius 12 at sixteen PRBs.
At this bandwidth, radii 12 and 24 require 2.906 and 2.995 GFLOPs per sample, respectively, with the same parameter count. We use radius 12 as the fixed default to obtain the demonstrated bandwidth-transfer performance at the lower forward-pass cost, while reporting radius 24 as a separate performance--cost option.

\begin{figure}[!htb]
    \centering
    \includegraphics[width=\columnwidth]{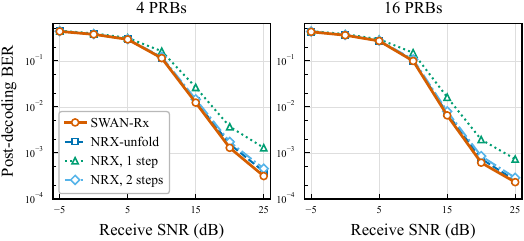}
    \caption{Post-decoding BER of one-, two-, and three-step NRX and \ourmethods{} on UMa: 4 PRBs (left); 16 PRBs (right).}
    \label{fig:nrx_budget}
\end{figure}
 \subsubsection{Lower BER and FLOPs Across NRX Depths}
The main comparisons use three-step NRX. We additionally evaluate one- and two-step variants to determine whether the reception--computation advantage of \ourmethods{} also holds against shallower, lower-cost NRX receivers. All three NRX models are trained separately from scratch with the same 4-PRB training sample budget, then compared with the fixed \ourmethods{} receiver at 4 and 16 PRBs without fine-tuning.

Reducing NRX depth lowers the parameter count and FLOPs in Table~\ref{tab:computational_cost}, but increases post-decoding BER in the evaluated runs (Fig.~\ref{fig:nrx_budget}). At both allocations, \ourmethods{} achieves lower post-decoding BER and fewer forward-pass operations than all three depth variants. In particular, one-step NRX has a similar cost, 0.737 versus 0.726 GFLOPs at 4 PRBs and 2.949 versus 2.906 GFLOPs at 16 PRBs, but higher BER. Thus, the reception advantage also holds at a comparable operation count, although one- and two-step NRX use fewer parameters.

\begin{figure}[!htb]
    \centering
    \includegraphics[width=\columnwidth]{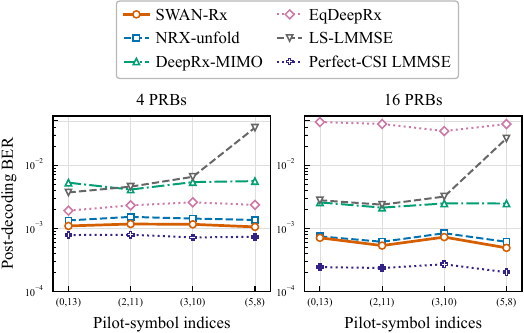}
    \caption{Post-decoding BER across 4 pilot layouts on UMa at 20~dB: 4 PRBs (left); 16 PRBs (right).}
    \label{fig:pilot_position}
\end{figure}
 \subsection{Pilot Analysis and Channel-Estimation Extension}
\label{ssec:pilot_analysis_extension}
\subsubsection{Shared Weights Support Multiple Pilot Layouts}
\label{ssec:icl_analysis}
We examine the BER of each pilot layout separately at 20~dB and allocations of 4 and 16 PRBs. The lower post-decoding BER of \ourmethods{} relative to NRX holds for every tested layout at both bandwidths (Fig.~\ref{fig:pilot_position}). All four layouts occur during training, so this result characterizes reception across known layouts.

\begin{figure}[!htb]
    \centering
    \includegraphics[width=\columnwidth]{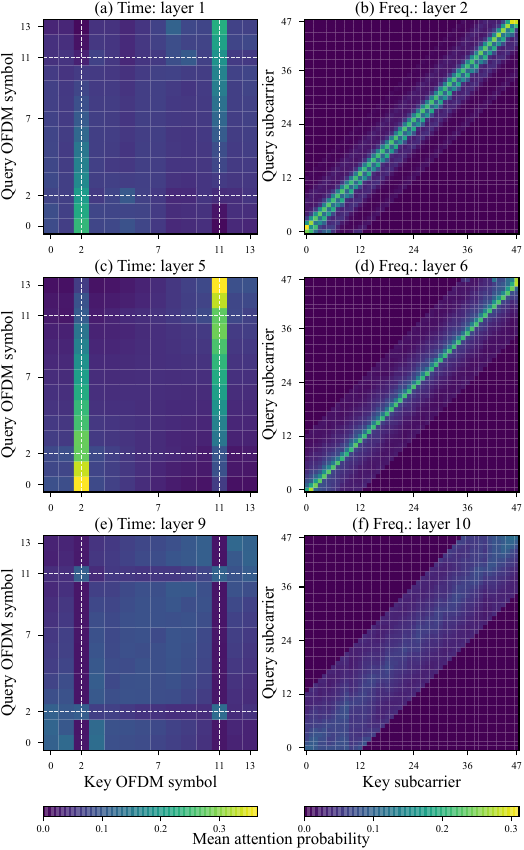}
    \caption{\ourmethods{} attention on UMa at 15~dB, 4 PRBs, and pilot symbols $(2,11)$. Left: time attention; right: frequency attention. White dashed lines mark pilot symbols; color scales are shared within each column.}
    \label{fig:attn}
\end{figure}
 We inspect attention from the frozen \ourmethods{} checkpoint on one predetermined realization under the conditions in Fig.~\ref{fig:attn}. From top to bottom, the maps show time-axis layers 1, 5, and 9 on the left and frequency-axis layers 2, 6, and 10 on the right. Each map averages the four heads and the orthogonal grid axis of the same frame. The frequency support is imposed by the radius-12 mask. This descriptive example does not isolate the causal contribution of pilot information.

\begin{figure}[!htb]
    \centering
    \includegraphics[width=\columnwidth]{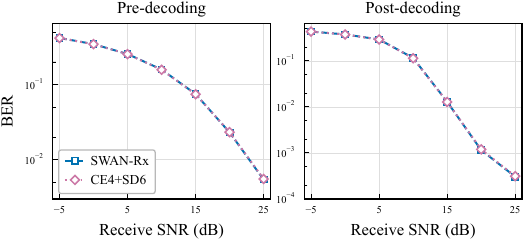}
    \caption{Detection BER of \ourmethods{} and CE4+SD6 on UMa at 4 PRBs. Both use QK normalization and SwiGLU. Left: pre-decoding BER; right: post-decoding BER.}
    \label{fig:ce_extension}
\end{figure}
 \subsubsection{Explicit Channel Output with Comparable Detection BER}
\label{ssec:ce_extension}
We finally examine the detection performance and channel-estimation accuracy of the optional CE4+SD6 extension. Both compared neural models use QK normalization and SwiGLU and the same four-PRB training budget. CE4+SD6 adds intermediate channel supervision and changes the stage allocation and frequency windows, so the BER comparison evaluates the complete extension rather than isolating the CE loss.

CE4+SD6 provides an explicit channel output with BER close to that of the detection-only variant over the evaluated SNR grid (Fig.~\ref{fig:ce_extension}). At 25~dB, their post-decoding BERs are $3.12\times10^{-4}$ and $3.05\times10^{-4}$, respectively.

\begin{figure}[!htb]
    \centering
    \includegraphics[width=\columnwidth]{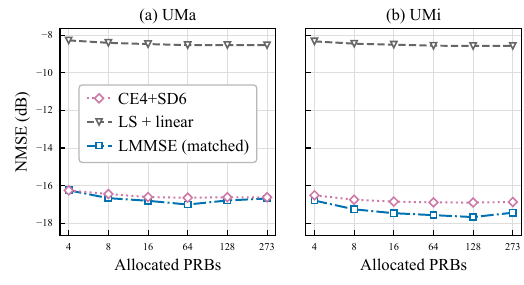}
    \caption{Channel-estimation NMSE on data REs at 10 dB: (a) UMa; (b) UMi. Each curve averages 4 pilot layouts. LMMSE uses independently calibrated covariance matched to each channel and bandwidth; CE4+SD6 uses frozen 4-PRB UMa-trained weights.}
    \label{fig:ce_nmse}
\end{figure}
 We assess the channel output against LS with linear interpolation and a pilot-based, two-dimensional LMMSE channel estimator on UMa and UMi at 10~dB. The ``LMMSE (matched)'' reference uses the known per-frame noise variance and a separable time--frequency covariance estimated for each channel and bandwidth from 4096 independent true-channel calibration frames, excluding all test frames. CE4+SD6 uses the same frozen 4-PRB UMa-trained weights throughout, without calibration on the target channel or bandwidth. Each channel, bandwidth, SNR, and pilot-layout condition uses 4096 paired test frames. We compute normalized mean-square error (NMSE) as the ratio of accumulated squared channel-estimation error to accumulated true-channel energy over data REs and all antenna pairs, then average the 4 pilot-layout values in linear units before conversion to dB.

On UMa, CE4+SD6 achieves NMSE between $-16.26$ and $-16.64$~dB over 4--273 PRBs (Fig.~\ref{fig:ce_nmse}(a)). It improves on LS with linear interpolation by approximately 8~dB and remains within 0.4~dB of the LMMSE reference with matched statistics. On UMi, its NMSE ranges from $-16.51$ to $-16.90$~dB, improving on LS by 8.2--8.4~dB and remaining within 0.8~dB of this reference (Fig.~\ref{fig:ce_nmse}(b)). The stable NMSE under both bandwidth expansion and joint channel--bandwidth transfer extends the evidence for frozen-weight reuse in Section~\ref{ssec:method_complexity} to the explicit channel output. Additional measurements on UMa at 20~dB give CE4+SD6 NMSE from $-20.72$ to $-21.02$~dB, again below LS, although LMMSE gives lower NMSE at every tested bandwidth.

\FloatBarrier
\section{Conclusion}
\label{sec:conclusion}

We presented \ourmethods{}, a MIMO-OFDM soft-detection receiver that combines a per-RE representation of pilot context and data observations with windowed axial attention. Shared mappings and relative frequency positions support variable allocations without bandwidth-specific learned parameters, while computation grows linearly with the subcarrier count for fixed symbol count, model dimensions, depth, and window radius. An optional CE+SD extension provides explicit channel estimates through intermediate supervision.

Extensive experiments demonstrated frozen-weight reuse from four-PRB training to allocations as large as 273 PRBs, together with transfer to the evaluated unseen statistical and ray-tracing channels.
At the four-PRB training bandwidth, the proposed receiver achieved lower pre- and post-decoding BER than the implemented three-step NRX and DeepRx-MIMO baselines over the evaluated SNR grid.
Under the common forward-pass accounting convention, it reduced FLOPs by 56.3\% and 35.7\%, respectively, while also using fewer parameters.
The selected optimized implementations also achieved 40.2--60.5\% lower batch-size-1 latency than NRX and 12.4--53.3\% lower latency than DeepRx-MIMO over 4--273 PRBs on the measured GPU.

\bibliographystyle{styles/IEEEtran}
\bibliography{refs}

\end{document}